\documentclass[aps,preprint,floats,epsf,epsfig,nofootinbib,letter]{revtex4}

\usepackage{graphicx}
\usepackage{dcolumn}
\usepackage{bm}
\usepackage{subfigure}
\usepackage{mathrsfs}

\def\be{\begin{eqnarray}}
\def\en{\end{eqnarray}}
\def\non{\nonumber}
\def\la{\langle}
\def\ra{\rangle}

\newcommand{\cov}[1]{\hspace{0.5mm}/\hspace{-2.6mm}#1}
\begin{document}

\renewcommand{\baselinestretch}{1.10}

\font\el=cmbx10 scaled \magstep2{\obeylines\hfill Aug., 2013}

\vskip 1.5 cm


\centerline{\Large\bf   A Study of Dirac
Fermionic Dark Matters}

\bigskip
\centerline{\bf Chun-Khiang Chua, Ron-Chou Hsieh}
\medskip
\centerline{Department of Physics and Chung Yuan Center for High Energy Physics,} 
\centerline{Chung Yuan Christian University,}
\centerline{Chung-Li, Taiwan 320, Republic of China}
\medskip

\centerline{\bf Abstract} We study pure weak eigenstate Dirac
fermionic dark matters (DM). We consider WIMP with renormalizable
interaction. According to results of direct searches and the nature
of DM (electrical neutral and being a pure weak eigenstate), the
quantum number of DM is determined to be $I_3=Y=0$. There are only
two possible cases: either DM has non-vanishing weak isospin ($I\neq
0$) or it is an isosinglet ($I=0$). In the first case, the
Sommerfeld enhancement is sizable for large $I$, producing large
$\chi^0\overline{\chi^0}\to VV$ rates. In particular, we obtain
large $\chi\bar\chi\to W^+ W^-$  cross section, which is comparable
to the latest bounds from indirect searches and $m_\chi$ is
constrained to be larger than few hundred GeV to few TeV. It is
possible to give correct relic density with $m_\chi$ higher than
these lower bounds. In the second case, to couple DM to standard
model (SM) particles, a SM-singlet vector mediator $X$ is required
from renormalizability and the SM gauge quantum numbers. To satisfy
the latest bounds of direct searches and to reproduce the DM relic
density at the same time, resonant enhancement in DM annihilation
diagram is needed. Thus, the masses of DM and the mediator are
related. Furthermore, this model is
not sufficient to explain the deviation in muon $g-2$.
\bigskip
\small

\pacs{Valid PACS appear here}

\maketitle

%
%

\section{Introduction}
It is known that the discrepancy in speed of galaxy in our universe
between observation and prediction from Newtonian gravitation theory
indicate that there must be something ``dark" there. These so called
dark matter (DM), according to the observation of Wilkinson
Microwave Anisotropy Probe (WMAP) and Planck, supply about 23$\%$ of
composition to our universe~\cite{WMAP,Planck}. Dark matter cannot
be observed from measuring their luminosity. Then, would it be
possible that they are something like black hole, neutral star,
brown dwarf, etc., which can only emit little or even no
electro-magnetic radiation. Big-Band nucleosynthesis (BBN) provides
powerful constraints on this account. From predictions of the
abundances of the light elements, D, $^3$He, $^4$He, etc., one can
evaluate the value of relic blackbody photon density as $\eta\equiv
n_\mathrm{b}/n_\gamma\approx(5.1-6.5)\times10^{-10}~$\cite{PDG}. The
measurements can be converted to the baryonic fraction of critical
density,
$\Omega_\mathrm{b}=\rho_\mathrm{b}/\rho_{\mathrm{crit}}\simeq(0.019-0.04)h^{-2}$,
where $h=0.72\pm0.08$ is the present Hubble parameter. The resulting
baryonic fraction $\Omega_\mathrm{b}$ is much smaller than the
latest result on cold DM fraction,
$\Omega_{\mathrm{DM}}h^2=0.1187\pm0.0017$~\cite{Planck}.
It tells us that, in the standard
model (SM) of particle physics, there is no candidate for DM.
Therefore, one has to extend the SM to account for the DM.

To construct a DM model, there are some basic requirements on DM.
DM must be stable, charge neutral
and have non-negligible
mass. ``Stable" means that it should live long enough that we can
still observe their relic. ``Neutral" is to avoid DM to shine and
``non-negligible mass" means that the DM can gather gravitationally
on small scales and so seed galaxy formation. There are many DM
candidates such as weakly interactive massive particles (WIMP),
axions, Kaluza-Klein mode in extra dimensions, etc..
For a
recent review of dark matter, see \cite{DG:12a}.

In this study, we will only consider the WIMP scenario.
DM only interact through the gravity and weakly
interacting force with interaction cross-sections basically not higher than the
weak scale. We investigate a renormalizable DM model by introducing a
pure weak eigenstate Dirac fermion as a DM candidate. We do not
consider scalar or Majorana DM, which have been discussed intensively in the
literature (see, for example,~\cite{Mc:94,BPV:01,HLLTT:09,GLPR:10,DGW:12,CTTYZ:12a,OY:13a}).
There are some Dirac fermionic DM models being considered in past
years~\cite{Zeldovich:1980st,KLS:08a,CY:07a,FLNP:10a,Andreas:2011in,Belotsky:2004st}.
For instance, fermionic DM contributing to indirect precesses~\cite{Zeldovich:1980st,Belotsky:2004st}, fermionic DMs with a charged scalar particle as
a mediator to couple to SM particles through renormalizable terms
are discussed in~\cite{KLS:08a}, while some use vector bosons, such
as $Z'$, to mediate interactions with SM
particles~\cite{CY:07a,FLNP:10a,Andreas:2011in}.
In this work the models we considerd are viewed as purely low energy models. 
The UV completion of the models is beyond the scope of the present work.

The lay out of this work is as following.
In the next section, 
we introduce a weak eigenstate Dirac ferminoic DM model with renormalizable
interaction. We try to develop the model logically with a bottom-up approah.
We then constrain the model using relic density, direct and indirect detection
experiments.
Numerical results are presented in Sec. III, which follows
by discussion and conclusion in Sec. IV. Some formulas are collected
in the Appendix.

\section{Framework}

In WIMP scenario 
one can
write down a simple DM model by adding on SM a single
Dirac fermionic multiplet $\chi$ with the Lagrangian such
as:~\cite{MDM}
\begin{eqnarray}
\mathscr{L} = \mathscr{L}_{\rm SM}
+ \bar{\chi} (i \cov{D}-m_\chi) {\chi},
\label{eq:L}
\end{eqnarray}
where the covariant derivative $D_\mu$ contains the known electroweak
gauge couplings to the vector bosons of the SM such that
\begin{eqnarray}\label{eq:covD}
D_\mu &=& \partial_\mu + i\frac{g}{\sqrt{2}}(W_\mu^{+}T^{+}
+W_\mu^{-}T^{-}) + i\frac{1}{\sqrt{g^2+{g'}^2}}Z_\mu(g^2T^3 -
{g'}^2Y) +i\frac{gg'}{\sqrt{g^2+{g'}^2}}A_\mu Q\non\\ &=&
\partial_\mu + i\frac{g}{\sqrt{2}}(W_\mu^{+}T^{+} +W_\mu^{-}T^{-}) +
i\frac{g}{\cos\theta_W}Z_\mu T^3.
\end{eqnarray}
Here, in the second line we have used the condition ``electric
charge neutrality", $Q=T^3+Y=0$, and the definition of weak mixing
angle, $\cos\theta_W=g/\sqrt{g^2+{g'}^2}$.
Note that in this work we only consider renomalizable interactions.
Therefore, the DM cannot couple to Higgs.
Furthermore, we may assign some $Z_2$ symmetry to maintain the stability of the DM.

In the Lagrangian,
the $Z$ boson interaction term will
produce a tree-level spin independent elastic cross sections with a
nucleus $N$:
\begin{equation}
\sigma^{\mathrm{SI}}_A(\chi N\rightarrow\chi N) =
\frac{{\mu_N}^2}{4\pi} {\left(\frac{g}{\cos\theta_W M_Z}\right)}^4
{I_3}^2 {\left[-\frac{1}{4}(A-Z) +
(\frac{1}{4}-{\sin\theta_W}^2)Z\right]}^2,
\end{equation}
where 
$Z$ and $A$
are the number of protons and of nucleons in the target nucleus,
$I_3$ is the weak isospin quantum number
and $\mu_N$ is the reduce mass of DM and nucleus. The above formula
gives a normalized cross section (see Appendix \ref{app:Formula})
\begin{equation}
\sigma _N^Z\simeq  I^2_3 \times 10^{-40}{\rm cm}^2,
\end{equation}
for $m_\chi$ ranges from few GeV to few TeV.
~\footnote{The mass of the fermionic DM should be
larger than GeV, which is known as the
Lee-Weinberg limit~\cite{LW:77a}.}
Therefore, the
magnitude of the cross section exceeds most of the experimental upper
bounds which obtained from direct detection searches
for 
$m_\chi\gtrsim10$ GeV~\cite{DMtools}.\footnote{In fact, the case of DM with
non-vanishing $T_3$ is still allowable for light WIMP candidates by
only consider the constraint from direct detection searches. There
are also some efforts are devoted to searching for DM with mass of
order $\lesssim10$ GeV~\cite{Hooper:12a}. But, here we do not
consider light DM case.}
The situation forces us to consider two
cases of heavy DM with different quantum numbers: (i) $I\neq0, I_3=Y=0$, and
(ii) $I=Y=0$.



Before we proceed to these two cases, it will be useful to recall some basics formulas.
To obtain the thermal relic density for DM, we must
solve the Boltzmann equation, which control the evolution of the DM
abundance,
\begin{equation}\label{Boltzmann eq}
\frac{dn_\chi}{dt} + 3Hn_\chi = -\langle\sigma_\mathrm{ann} v\rangle_{\chi\bar\chi}
[n_\chi n_{\bar{\chi}} - n_\chi^\mathrm{eq}
n_{\bar{\chi}}^\mathrm{eq}],
\end{equation}
where $H\equiv
\dot{a}/a=\sqrt{4\pi^3g_*(T)T^4/(45M_{\mathrm{PL}}{}^2)}$ is the
Hubble parameter, $M_{\mathrm{PL}}$ is the Planck mass,
$g_*$ is the total effective numbers of relativistic degrees of
freedom~\cite{KTbook,CR:03}, $\langle\sigma_\mathrm{ann} v\rangle_{\chi\bar\chi}$ is the thermal averaged 
$\chi\bar\chi$ annihilation cross section and $n_\chi(n_{\bar{\chi}})$ is the
number density of DM (anti-DM). Note that for the Dirac fermionic DM, we have
$n_{DM}=n_\chi+n_{\bar \chi}=2 n_{\chi(\bar\chi)}$ for the DM number density, and, consequently, we obtain
\begin{equation}\label{Boltzmann eq1}
\frac{dn_{DM}}{dt} + 3Hn_{DM} = -\frac{\langle\sigma_\mathrm{ann} v\rangle_{\chi\bar\chi}}{2}
\left[n^2_{DM} - (n_{DM}^\mathrm{eq})^2\right]
=-\langle\sigma_\mathrm{ann} v\rangle
\left[n^2_{DM} - (n_{DM}^\mathrm{eq})^2\right],
\end{equation}
where we define $\langle\sigma_{\mathrm{ann}}v\rangle
\equiv{\langle\sigma_{\mathrm{ann}}v\rangle_{\chi\bar\chi}}/{2}$, 
such that the Boltzmann equation can take the usual form.
The physical reasoning of the factor $1/2$ can be understood as following. 
DM can be separated into two halves ($\chi$ and $\bar\chi$). Half of the DM ($\chi$ or $\bar\chi$) can only annihilate with the other half of the DM ($\bar\chi$ or $\chi$) and vice versa, giving factor $1/4$ each. 
Therefore, by adding these two halves, we obtain the factor $1/2$.
We note in passing that $DM+N$ elastic scattering cross section, does not need the factor $1/2$, since it is compansated 
by a factor 2 arisen from $\chi+N$ and $\bar\chi+N$ scatterings. 

Following the standard procedure~\cite{KTbook} to solve
Eq.(\ref{Boltzmann eq}) approximately, we obtain the relations:
\begin{equation}\label{abundance}
\Omega _{\text{DM}}h^2 \approx 1.04\times
10^9\frac{{\rm GeV}^{-1}}{M_{\mathrm{PL}} \sqrt{g_*\left(T_f\right)}J(x_f)},
\end{equation}
\begin{equation}\label{xf}
x_f \approx \ln\left[\frac{2\times0.038 m_\chi
M_{\mathrm{PL}}\langle\sigma_{\mathrm{ann}}v\rangle
}{\sqrt{g_*\left(T_f\right)}{}x_f^{1/2}}\right],
\end{equation}
where we have
\begin{equation}\label{Jfactor}
J\left(x_f\right) \equiv \int_{x_f}^{\infty }
\frac{\langle\sigma_{\mathrm{ann}}v\rangle}{x^2} \, dx
\end{equation}
with $x_f$ defined as $m_\chi/T_f$ and $T_f$ being
the freeze-out temperature,
and the thermal averaged annihilation cross section $\langle\sigma_{\mathrm{ann}}v\rangle$
with  $v$ the ``relative velocity'' is defined as
\begin{eqnarray}\label{T_average}
\langle\sigma_{\mathrm{ann}}v\rangle
\equiv\frac{\langle\sigma_{\mathrm{ann}}v\rangle_{\chi\bar\chi}}{2}
&\equiv&\frac{3\sqrt 6}{\sqrt{\pi}v_0^3}\int_0^\infty dv\, v^2
\frac{(\sigma_{\mathrm{ann}}v)_{\chi\bar\chi}}{2}
e^{-3v^2/2v_0^2}
\non\\
&=&
\frac{x^{3/2}}{2\sqrt{\pi}}\int_0^\infty dv\, v^2
\frac{(\sigma_{\mathrm{ann}}v)_{\chi\bar\chi}}{2}e^{-xv^2/4},
\end{eqnarray}
where we define $v_0\equiv\la v^2\ra^{1/2}$ and $v_0=\sqrt{6/x_f}$ has been used in the last expression.
It is straightforward to obtain
\be
J\left(x_f\right) = \int_{x_f}^{\infty }
\frac{\langle\sigma_{\mathrm{ann}}v\rangle}{x^2} \, dx
=\int_0^\infty dv\frac{(\sigma v)_{\chi\bar\chi}}{2} v\left[1-{\rm erf}\left(v\sqrt x_f/2\right)\right].
\en
Note that in the above equations factors of $1/2$ arisen from $n_{DM}=n_\chi+n_{\bar\chi}=2n_{\chi(\bar\chi)}$ are included.
We can now turn to the formalisms for the two above mentioned cases.



\begin{figure}[t] \centering
\includegraphics[width=0.8\textwidth]{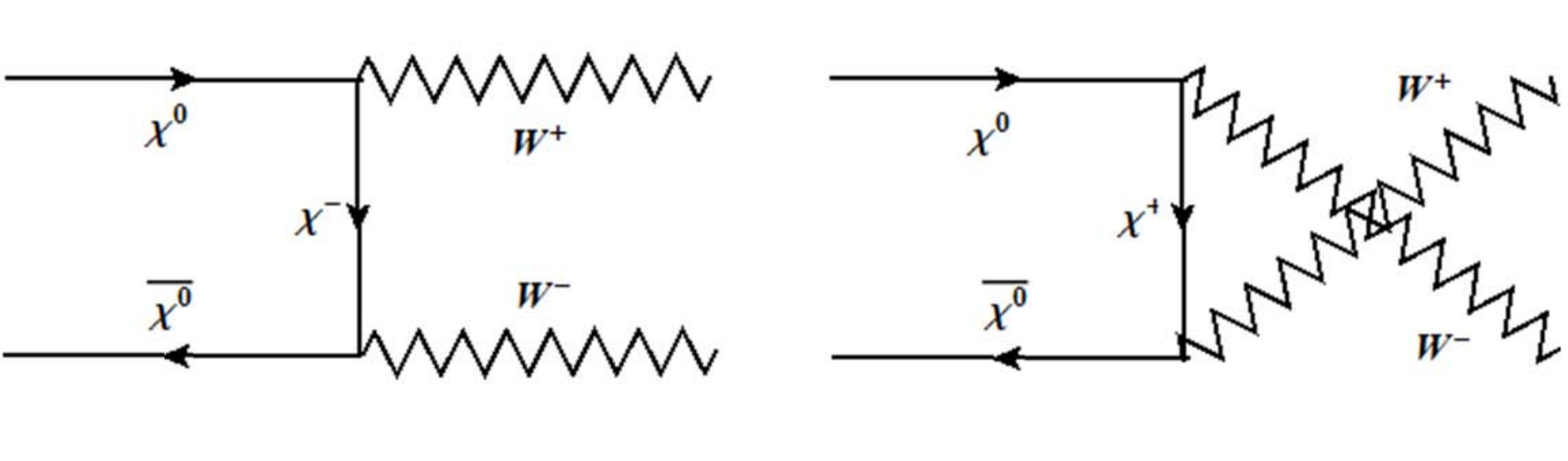}
\caption{The Feynman diagrams of DM annihilation for $W^+W^-$
channel.} \label{Fdiagram}
\end{figure}

\subsection*{I. $\mathbf{I\neq0, I_3=Y=0}$ case}

In this case, the DM possesses non-vanishing weak isospin $I$ but
with zero hypercharge. The constraint condition, $I_3=0$, indeed
avoids the troublesome $Z$ diagram. 
However, the
contribution from the $W$ boson interaction needs to be investigated as well. 
Note that this case was also studied in \cite{MDM,Cirelli:2009uv}.
For completeness, we shall include them in this analysis.
In fact, this work differs from the previous studies in several aspects.
We focus on the Dirac Fermionic DM case. 
We are interested in finding the direct consequences of Eq.~(\ref{eq:L}) instead of completing the model by adding other ingreedients.
Therefore, all isospin assignments are kept.
As we will discuss later, the Sommerfeld effects applicable to any isospin assignment will also be given.
Furthermore, we are in a position that new data, such as galactic annihilation rate~\cite{FermiLat:11a}, is available and can be compared to.

The DM pair
can annihilate into a $W$ boson pair (see Fig.~\ref{Fdiagram}) and
then can contribute to
the relic density of DM and indirect processes from milky
way satellites. 
The $\chi^0\overline{\chi^0}\to W^+W^-$ annihilation cross section
contributed from Fig.~1 for case I is calculated to be
\begin{eqnarray}
(\sigma_{\mathrm{ann}}v)_{\chi\bar\chi} &=& [I(I+1)]^2\frac{ g^4 \sqrt{s-4 m_W^2}
}{32\pi s^{3/2} \left(s-2 m_W^2\right)}\left\{\frac{\left(2
m_W^2-s\right)\left(s m_{\chi }^2+4 m_{\chi }^4+2
m_W^4\right)}{\left(m_{\chi }^2 \left(s-4
m_W^2\right)+m_W^4\right)}\right.\non\\ &+& \left.\frac{ \left(4
m_{\chi }^2 \left(s-2 m_W^2\right)-8 m_{\chi }^4+4
m_W^4+s^2\right)}{\sqrt{\left(s-4 m_{\chi }^2\right) \left(s-4
m_W^2\right)}} \log \left[-\frac{\sqrt{\left(s-4 m_{\chi }^2\right)
\left(s-4 m_W^2\right)}-2 m_W^2+s}{\sqrt{\left(s-4 m_{\chi
}^2\right) \left(s-4 m_W^2\right)}+2 m_W^2-s}\right]\right\}.\non\\
\label{eq:sigmav xx->WW}
\end{eqnarray}
After substituting $s=4m_\chi^2+m_\chi^2 v^2$ into the above equation and expanding around $v^2$, one obtains:
\begin{eqnarray}
\langle\sigma_{\mathrm{ann}}v\rangle = \langle a^{+-} + b^{+-} v^2 + {\cal
O}(v^4)\rangle,
\label{eq:sigmavann}
\end{eqnarray}
where we have
\begin{eqnarray}
a^{+-} &\equiv& [I(I+1)]^2 \frac{g^4(m_\chi^2-m_W^2)^{3/2}}
{16\pi m_\chi (2m_\chi^2-m_W^2)^2},
\non\\
b^{+-} &\equiv&  [I(I+1)]^2\frac{g^4 (m_{\chi }^2-m_W^2)^{1/2}
\left(24 m_{\chi }^6+28 m_{\chi }^4m_W^2
-36 m_{\chi }^2 m_W^4+17 m_W^6\right)}
{384 \pi m_{\chi }  \left(2m_{\chi }^2 -
m_W^2\right)^4},
\label{eq:ab}
\end{eqnarray}
with $g=e/\sin\theta_W$ and factor $1/2$ arisen from the Dirac DM are included. 
We
find that neglecting $v^4$ and higher order terms is a good
approximation. In fact, substituting $v^2=\la v^2\ra$
into $\sigma v$ almost gives identical results to the above
approximated results. 
For thermal relic abundance, we have $\langle v^2\rangle=6
x^{-1}_f$ from Eq.~(\ref{T_average}), and, consequently,
\be
\langle\sigma_{\mathrm{ann}}v\rangle \simeq  a^{+-} +6 \frac{b^{+-}}{ x_f},
\qquad
J(x_f)\simeq \frac{a^{+-} +3 b^{+-}/ x_f}{x_f}.
\en
Note that for $\langle\sigma_{\mathrm{ann}}v\rangle$ of indirect processes from milky
way satellites, we have the thermal average quantity $\langle
v^2_{1,2}\rangle=v_0^2/2$, where $v_0$ is chosen
to be the canonical value 270$\sqrt2$
km/s~\cite{JKG:96a}.

It is known that we need to take into account Sommerfeld enhancement
effect, when the velocity is very small~\cite{ArkaniHamed:2008qn}.~\footnote{Some
authors\cite{Belotsky:2004st} also called this as Sakharov
effect~\cite{Sakharov:1948yq}.} In the elastic scattering case, the
cross-section receives Sommerfeld enhancement as \be \sigma
v=(\sigma v)_0 S, \label{eq:elastic} \en where $(\sigma v)_0$
corresponds to the perturbative result and $S$ is the Sommerfeld
enhancement factor. Equivalently, the amplitude receives a $S^{1/2}$
factor. For a force carrier with mass $m_\phi$ and couplings
$\alpha$, the Sommerfeld factor is given by~\cite{Feng:2010zp} \be
S(\alpha)=\frac{\pi }{\epsilon_v}
\frac{\sinh\left(\frac{2\pi\epsilon_v}{\pi^2\epsilon_\phi/6}\right)}
{\cosh\left(\frac{2\pi\epsilon_v}{\pi^2\epsilon_\phi/6}\right)
-\cos\left(2\pi
\sqrt{\frac{1}{\pi^2\epsilon_\phi/6}-\frac{\epsilon^2_v}{(\pi^2\epsilon_\phi/6)^2}}\right)},
\label{eq:S} \en with \be \epsilon_v\equiv\frac{v}{\alpha}, \quad
\epsilon_\phi\equiv\frac{m_\phi}{\alpha m_\chi}. \en Note that we
have $S>1$ for $\alpha>0$ and vise verse.
%

\begin{figure}[t]
\centering
 \subfigure[]{
  \includegraphics[width=0.45\textwidth]{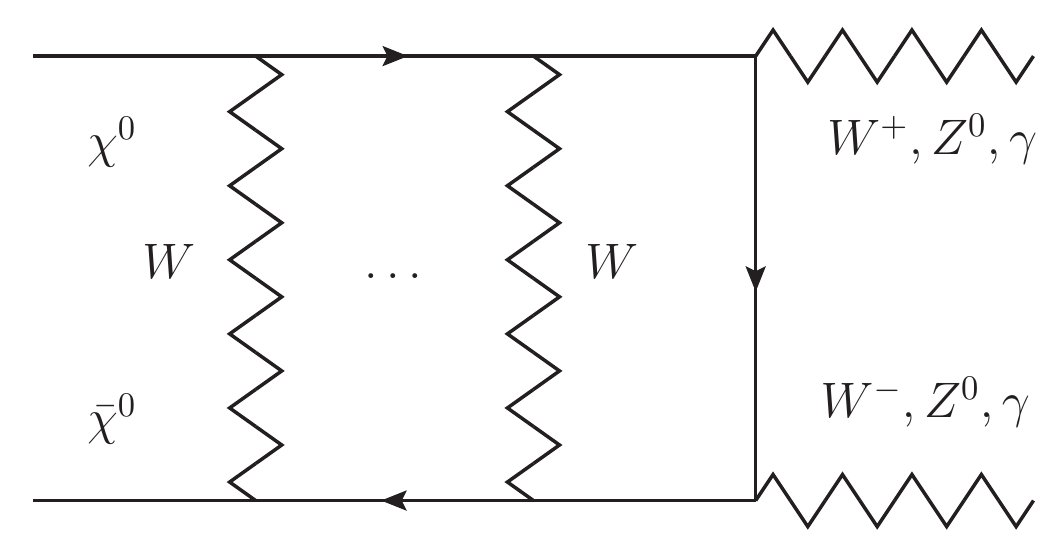}}
 \subfigure[]{
  \includegraphics[width=0.45\textwidth]{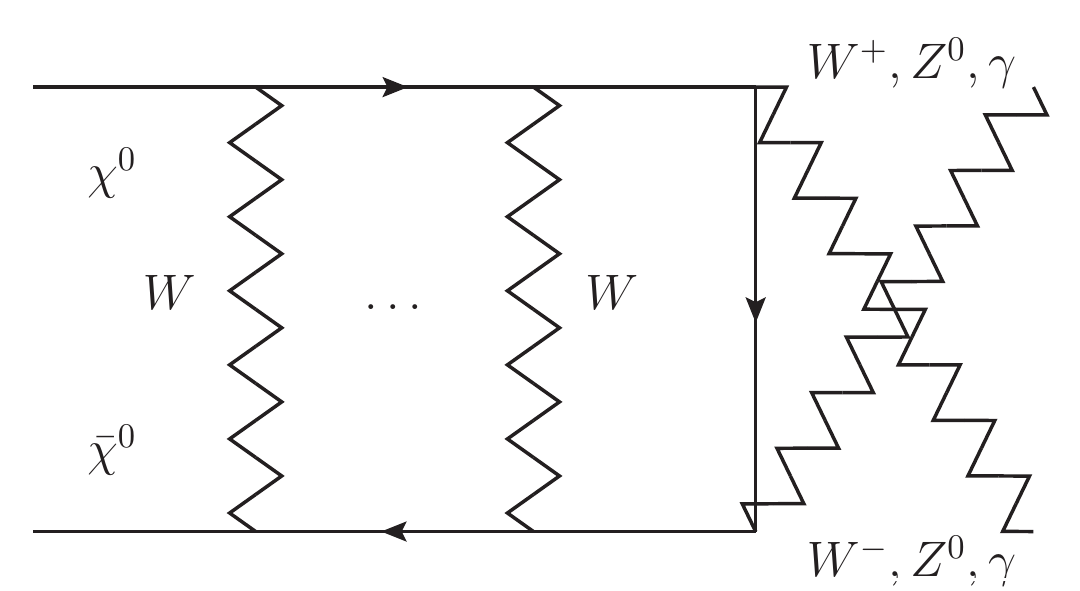}}
  \subfigure[]{
  \includegraphics[width=0.45\textwidth]{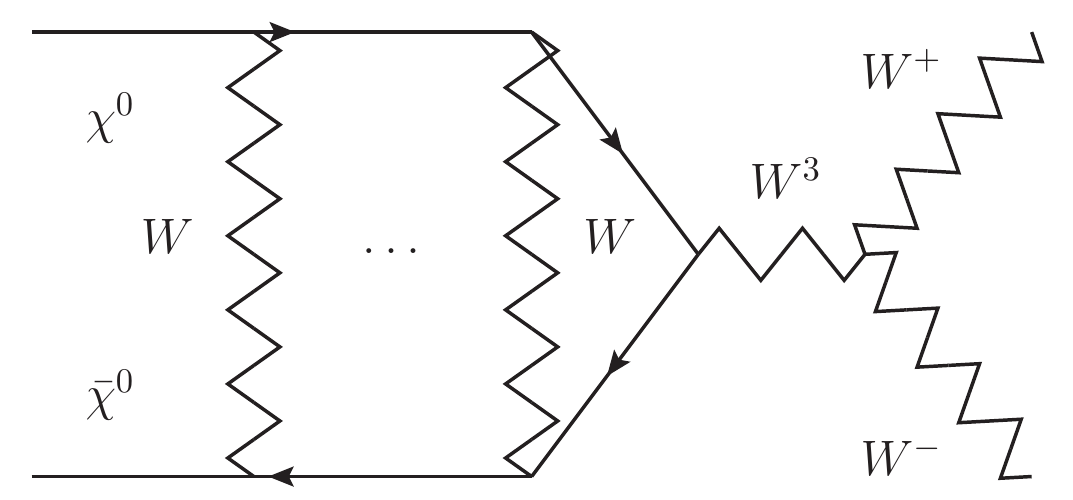}
}
\caption{(a) to (c):  $\chi^0\bar\chi^0\to VV$ annihilation diagrams with the Sommerfeld effect.}
\label{fig:Sommerfeld}
\end{figure}

The Sommerfeld enhancement in the present case is rather involved,
since the $\chi^0\overline{\chi^0}$ state can rescatter into other
states, such as $\chi^\pm\overline{\chi^{\pm}}$ and so on, through
$t$-channel diagrams by exchanging $W$ and $Z$ with the rescattered
state annihilated into $W^+W^-$ (see Fig.~\ref{fig:Sommerfeld}). 
To simplify the calculation we
follow \cite{Strumia:2008cf,Cirelli:2009uv} to consider the SU(2)
symmetric limit. For a generic isospin $I$, scatterings
$\chi^j\overline{\chi^j}\to \chi^i\overline{\chi^i}$ (with
$i,j=-I,-I+1,\dots,I-1,I$) produce a potential
$V_{ij}=-|V_W|\sum_{c=1,2,3}T^c_{ij}T^c_{ji}$
with $|V_W|=\alpha_W e^{-m_{W,Z} r}/r$.~\footnote{Note that we
differ from \cite{Strumia:2008cf,Cirelli:2009uv} as we do not
consider $\overline{\chi^i}$ to be identical to $\chi^{-i}$.
Therefore we do not have the factor of $\sqrt2$ on the
$\chi^0\overline{\chi^0}$ state (for $i$ or $j=0$) from the
identical particle effect and we have a $V$ matrix with larger
dimension.}

To proceed we use a procedure that is similar to those used in the study of final state interaction \cite{Chua:2007cm}.
We note that the potential can be diagonalized into several irreducible representations:
\footnote{The expression is obtained with the help of $\sum_c T^c_{ij} T^c_{ji}=-\sum_c T^c_{ij} T^c_{-i-j}(-)^{i-j}$ and the standard method of addition of angular momentum.}
\be
V_{ij}=\sum_{L=1}^{2I}(U^T)_{iL}\{-[I(I+1)-L(L+1)/2]|V_W|\}U_{Lj},
\en
with
\be
U_{Lj}=(-1)^j\la I j I(-j)|L0\ra,
\en
where $\la I j I(-j)|L0\ra$ is the Clebsch�Gordan coefficient (in the $\la j_1 m_1 j_2 m_2|JM\ra$ notation).
The irreducible parts of $V$ do not mixed in further rescattering as it is easy to see that
$(V^n)_{ij}=\sum_L U^T_{iL} \{-[I(I+1)-L(L+1)/2]|V_W|\}^n U_{Lj}$.
The Sommerfeld enhancement factor of the irreducible parts can be obtained as the elastic case and, consequently, we have
\be
{\cal S}_{ij}
=\sum_{L=1}^{2I} U^T_{iL}
S([I(I+1)-L(L+1)/2]\alpha_W)
U_{Lj},
\en
where $S(\alpha)$ is given by Eq.~(\ref{eq:S}) but with $m_\phi=m_{W,Z}$.
The $\chi^0\overline{\chi^0}\to W^+ W^-$ amplitude with Sommerfeld enhancement, $A_S$, is now given by
\be
A_S(\chi^0\overline{\chi^0}\to W^+W^-)&=&\sum_i A(\chi^i\overline{\chi^i}\to W^+W^-) {\cal S}^{1/2}_{i0},
\en
where $i$ is summed over all $\chi^i\overline{\chi^i}$ states. Therefore, the Sommerfeld enhanced $s$-wave part of $\sigma v$ is given by
\be
a^{+-}_S=\sum_{i,j} {\cal S}^{1/2}_{0i} a^{+-}_{ij} {\cal S}^{1/2}_{j 0},
\en
where $i$ and $j$ are summed over $\chi^i\overline{\chi^i}$ and $\chi^j\overline{\chi^j}$ states, repectively, and $a^{+-}_{ij}$ corresponds to the contribution from the $A^*(\chi^i\overline{\chi^i}\to W^+W^-) A(\chi^j\overline{\chi^j}\to W^+W^-)$ part.

It is straightforward to obtain
\be
a^{+-}_{ij}&=&\frac{g^4(m_\chi^2-m^2_W)^{3/2}}
{32\pi m_\chi(4m^2_\chi-m_W^2)^2(2m^2_\chi-m_W^2)^2}
\{2[I(I+1)-i^2][I(I+1)-j^2](4m^2_\chi-m_W^2)^2
\non\\
&&+ij(4m^3_\chi+20m^2_\chi m^2_W+3m^4_W)\}
\en
with factor $1/2$ included, and, consequently,
\be
a^{+-}_S=a^{+-}\,
\frac{1}{9}\left[2 S^{1/2}(I(I+1)\alpha_W)+S^{1/2}([-3+I(I+1)]\alpha_W)\right]^2.
\en
Note that a similar expression holds for the Sommerfeld enhanced $b$ term ($b_S$).
Finally, 
we obtain
\be
\la \sigma^{+-} v\ra=
\left
\la (\sigma^{+-} v)_0
\frac{1}{9}\left[2 S^{1/2}(I(I+1)\alpha_W)+S^{1/2}([-3+I(I+1)]\alpha_W)\right]^2
\right\ra.
\en
%
In the $S\to 1$ limit $\la\sigma^{+-} v\ra$ reduces to the one given in Eq.~(\ref{eq:sigmavann}). Furthermore, if the eigenvalues of $V_{ij}$ were degenerate, we return to the elastic result as in Eq.~(\ref{eq:elastic}).

Note that through rescattering we can also have $\chi^0\overline{\chi^0}\to Z^0Z^0, Z^0\gamma,\gamma\gamma$ annihilations, with
\be
A_S(\chi^0\overline{\chi^0}\to Z^0Z^0)&=&\sum_i A(\chi^i\overline{\chi^i}\to Z^0Z^0) {\cal S}^{1/2}_{i0}
\non\\
A_S(\chi^0\overline{\chi^0}\to Z^0\gamma)&=&\sum_i A(\chi^i\overline{\chi^i}\to Z^0\gamma) {\cal S}^{1/2}_{i0},
\non\\
A_S(\chi^0\overline{\chi^0}\to \gamma\gamma)&=&\sum_i A(\chi^i\overline{\chi^i}\to \gamma\gamma) {\cal S}^{1/2}_{i0},
\en
and, consequently,
\be
a^{00,0\gamma,\gamma\gamma}_S=\sum_{i,j} {\cal S}^{1/2}_{0i} a^{00,0\gamma,\gamma\gamma}_{ij} {\cal S}^{1/2}_{j 0},
\en
with
\be
a^{00}_{ij}=\frac{g^4\cos^4\theta_W(m_\chi^2-m^2_Z)^{3/2}i^2j^2}{8\pi m_\chi(2m^2_\chi-m_Z^2)^2},
\quad
a^{0\gamma}_{ij}=\frac{e^2g^2\cos^2\theta_W(4m_\chi^2-m^2_Z) i^2j^2}
{64\pi m^4_\chi},
\quad
a^{\gamma\gamma}_{ij}=\frac{e^4 i^2j^2}{32\pi m^2_\chi},
\en
and similar expressions for $b$ terms.
There processes also contribute to the relic density and are the inevitable consequnces and signatures of inelastic Sommerfeld effects.

We obtain the annihilation cross sections for $\chi^0\overline{\chi^0}\to Z^0Z^0,Z^0\gamma,\gamma\gamma$ as
\be
\la \sigma^{\alpha} v\ra=
\left
\la (a^{\alpha}+b^{\alpha} v^2)
\frac{1}{9}\left[S^{1/2}(I(I+1)\alpha_W)-S^{1/2}([-3+I(I+1)]\alpha_W)\right]^2
\right\ra,
\en
with $\alpha=00,0\gamma,\gamma\gamma$ and
\be
(a^{00},b^{00})&=&2(a^{+-},b^{+-})\big|_{g\to g\cos\theta_W,m_W\to m_Z},
\non\\
(a^{\gamma\gamma},b^{\gamma\gamma})&=&2(a^{+-},b^{+-})\big|_{g\to e, m_W\to 0},
\non\\
a^{0\gamma}&=&[I(I+1)]^2\frac{e^2g^2\cos^2\theta_W(4m_\chi^2-m^2_Z) }
{64\pi m^4_\chi},
\non\\
b^{0\gamma}&=&[I(I+1)]^2 e^2 g^2 \cos^2\theta\frac{12 m_\chi^4+13 m^2_\chi m_Z^2-m_Z^4}
{192\pi m_\chi^4 (4m_\chi^2-m_Z^2)},
\en
where factor $1/2$ are included.
It is clear that these $\la\sigma^\alpha v\ra$s go to zero in the $S\to 1$ or in the degenerate limit.
Note that we do not include loop contribution in these modes, since in most cases the contributions form inelastic rescattering parts are larger than the perturbative ones.
We are ready to perform numerical study, where results will be given in the next section.

\subsection*{II. $\mathbf{I=Y=0}$ case}

In this case, the DM candidate is a pure weak isospin singlet Dirac fermion. 
The case that
DM is a scalar has been discussed by others~\cite{Mc:94,HLLTT:09}.
To reproduce the observed relic density, we need to couple $\chi$ to
SM fermions $f$. We consider renormalizable interaction only.
Therefore, an additional particle $X$ is necessary to mediate the
$\chi\bar\chi\to f\bar f$ annihilation process. Since the DM is a
weak isospin singlet, the mediator can only be a singlet and the
$\bar f f$ bilinear term that couple to $X$ should be a singlet as
well. It is easy to see that the $\bar f f$ bilinear term can only
take the forms of $\bar f_L\gamma_\mu f_L$ and $\bar f_R\gamma_\mu
f_R$, and hence the mediator particle $X$ should be a vector
boson, if only renormalizable interaction is allowed.~\footnote{ We
are different from \cite{TY:13} in this respect, where they have
scalar mediator.}

The Lagrangian involving $\chi$, $f$ and $X$ is given by:
\begin{eqnarray}\label{DM_lagrangian}
\mathscr{L} &=& \mathscr{L}_{\rm SM} + \bar{\chi}
\left(i\cov{D}^{(\chi)}-m_\chi\right)\chi +\sum_f
\left(\bar{f_L}i\cov{D}^{(L)} f_L + \bar{f_R}i\cov{D}^{(R)} f_R -
\lambda_f\bar{f}_LH f_R -
\lambda_f\bar{f}_RH^\dagger f_L\right)\non\\ & &
-\frac{1}{4}\mathscr{X}^{\mu\nu}\mathscr{X}_{\mu\nu} +
\frac{1}{2}M_X^2 X^\mu X_\mu.
\end{eqnarray}
with
\begin{equation}
D_\mu^{(\chi)} \chi = \left(\partial_\mu + ig_\chi
X_\mu\right)\chi,\hspace{0.5in} D_\mu^{(L,R)} f_{(L,R)} =
\left(\partial_\mu + ig_f^{(L,R)} X_\mu\right)f_{(L,R)}
\end{equation}
and $\mathscr{X}_{\mu\nu} =\partial_\mu X_\nu-\partial_\nu X_\mu$ and the SM fermions $f$s pick up masses from Higgs mechanism (using the Higgs doublet $H$) as usual.
Here $g_\chi$, $g_f^{(L,R)}$ are corresponding coupling constants and $f_{(L,R)}$ is left (right) fermion.
For simplicity, we only consider a vector-type interaction, $g_f^L=g_f^R$.
The interaction term of the Lagrangian can be recast as:
\begin{equation}
\mathscr{L}_{\mathrm{int}} =- g_\chi\bar{\chi}\gamma^\mu \chi X_\mu
 -\sum_f g_f^V \bar{f}\gamma^\mu f X_\mu
\end{equation}
with $g_f^V=\frac{1}{2}(g_f^L+g_f^R)$. In order to determine the
relic density of DM 
particles, we need to calculate the cross section of DM annihilation
to fermion pairs. The result is given by
\begin{equation}\label{eq:sigma ann}
(\sigma_{\mathrm{ann}})_{\chi\bar\chi} = \frac{M_X}{\sqrt{s}}\times \frac{g_\chi^2
}{\left(s-M_X^2\right)^2+M_X^2 \Gamma
_{\text{tot}}^2}\frac{\sum_f\Gamma(\tilde{X}\rightarrow \bar{f}f)}
{\sqrt{s-4m_\chi^2}}\left(s+2m_\chi^2\right),
\end{equation}
with
\begin{equation}\label{decay}
\Gamma(\tilde{X}\rightarrow \bar{f}f) \equiv \frac{N_f^c
{g_f^V}^2\sqrt{{M_{\tilde{X}}}^2-4m_f^2}}{12\pi
{M_{\tilde{X}}}^2}\left({M_{\tilde{X}}}^2+2m_f^2\right),
\end{equation}
where $s=2m_\chi{}^2(1+1/\sqrt{1-v^2})$ is the square of the
center-of-mass energy; $\Gamma(\tilde{X}\rightarrow \bar{f}f)$ is
the decay width of ``virtual'' $X$ with mass
${M_{\tilde{X}}}=\sqrt{s}$ and $N_f^c$ is the number of color of the
$f$-fermion.

We have to calculate $\langle\sigma_{\mathrm{ann}}v\rangle$ 
numerically,
since the standard method (Taylor expand) gives extremely poor
results near the pole, even producing negative cross section
\cite{GS:91}.\footnote{It is noted that the original integrated
upper limit is infinity (see Eq.(\ref{T_average})). Here we modified
it to $1$, because the relative velocity $v$ cannot be larger than
light speed.} We can determine the validity parameter space of
$g_\chi$ and $g_f^V$ using the constraint from thermal relic
abundance and direct detection for any given values of $m_\chi$,
$M_X$. Note that in this case, the Sommerfeld enhancement is
irrelevant. As we shall see, we need to make use of the resonant
effect to give viable results on relic density without violating the
direct search data. In that region ($m_\chi\sim m_X/2$), the
Sommerfeld factor $S$ as given by Eq.~(\ref{eq:S}) is very close to
unity.

\section{Numerical Results}

\subsection*{I. $\mathbf{I\neq0, I_3=Y=0}$ case}

\begin{figure}[t]
\centering
 \subfigure[]{
  \includegraphics[width=0.45\textwidth]{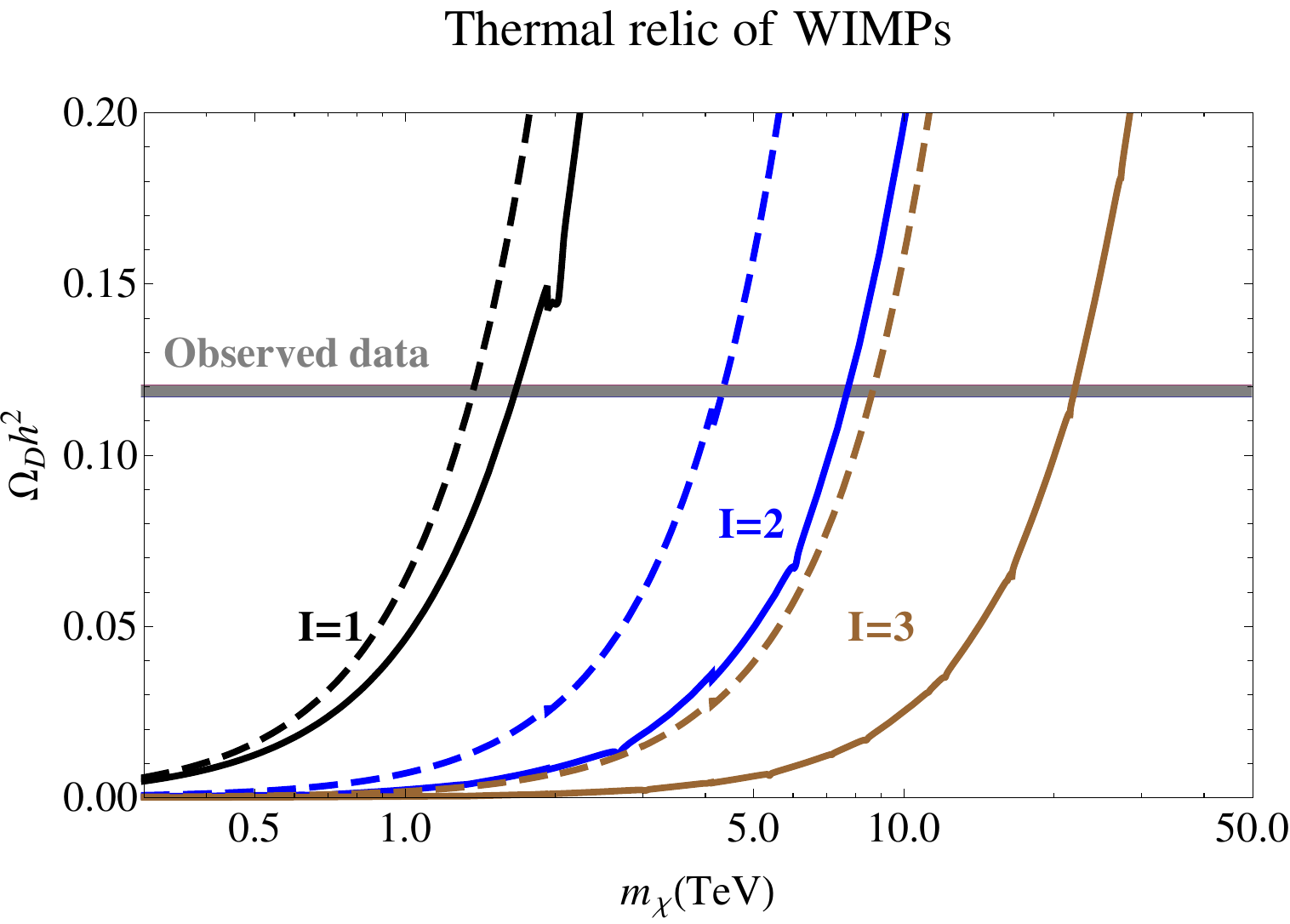}
}\subfigure[]{
  \includegraphics[width=0.45\textwidth]{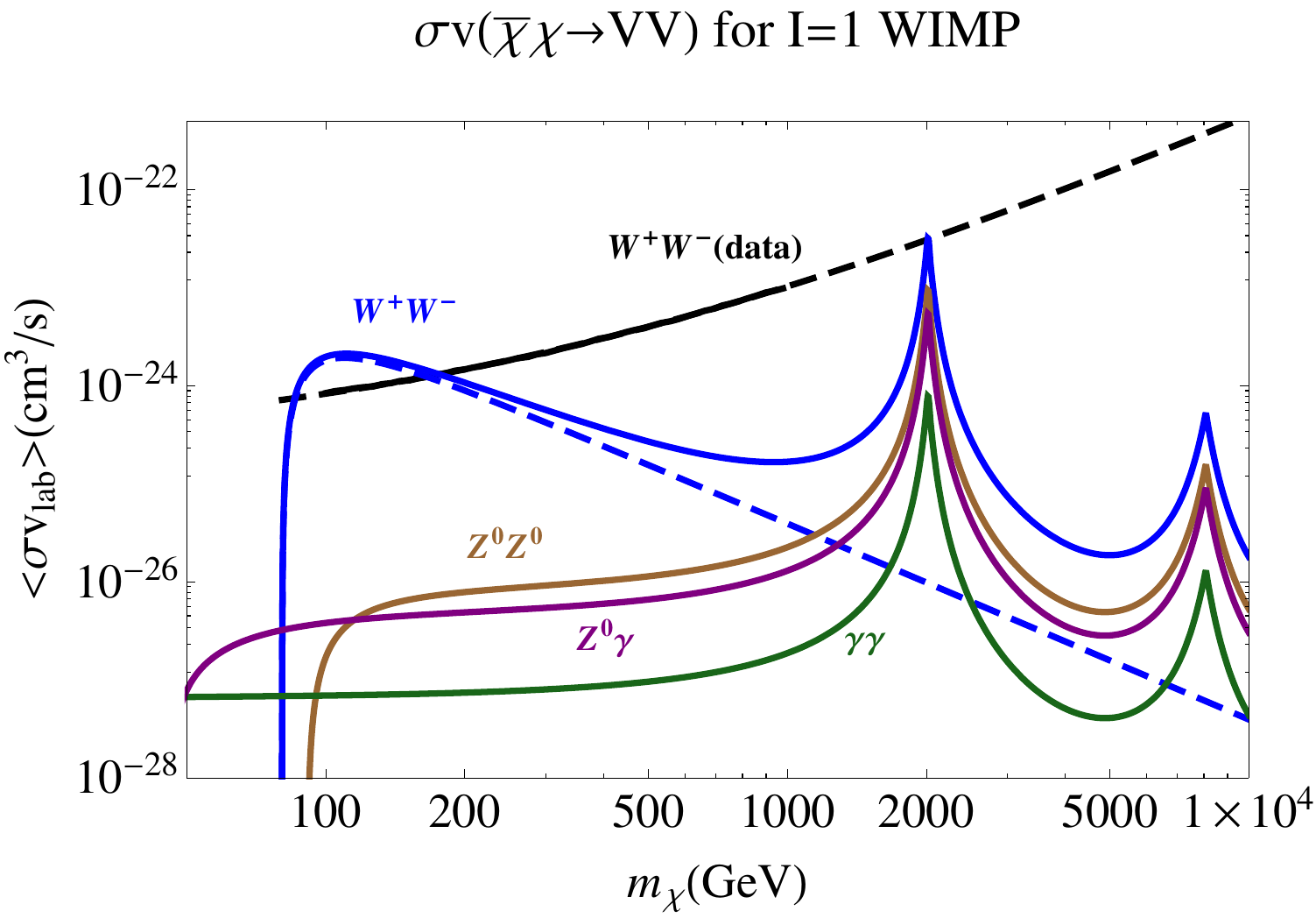}
}\\\subfigure[]{
  \includegraphics[width=0.45\textwidth]{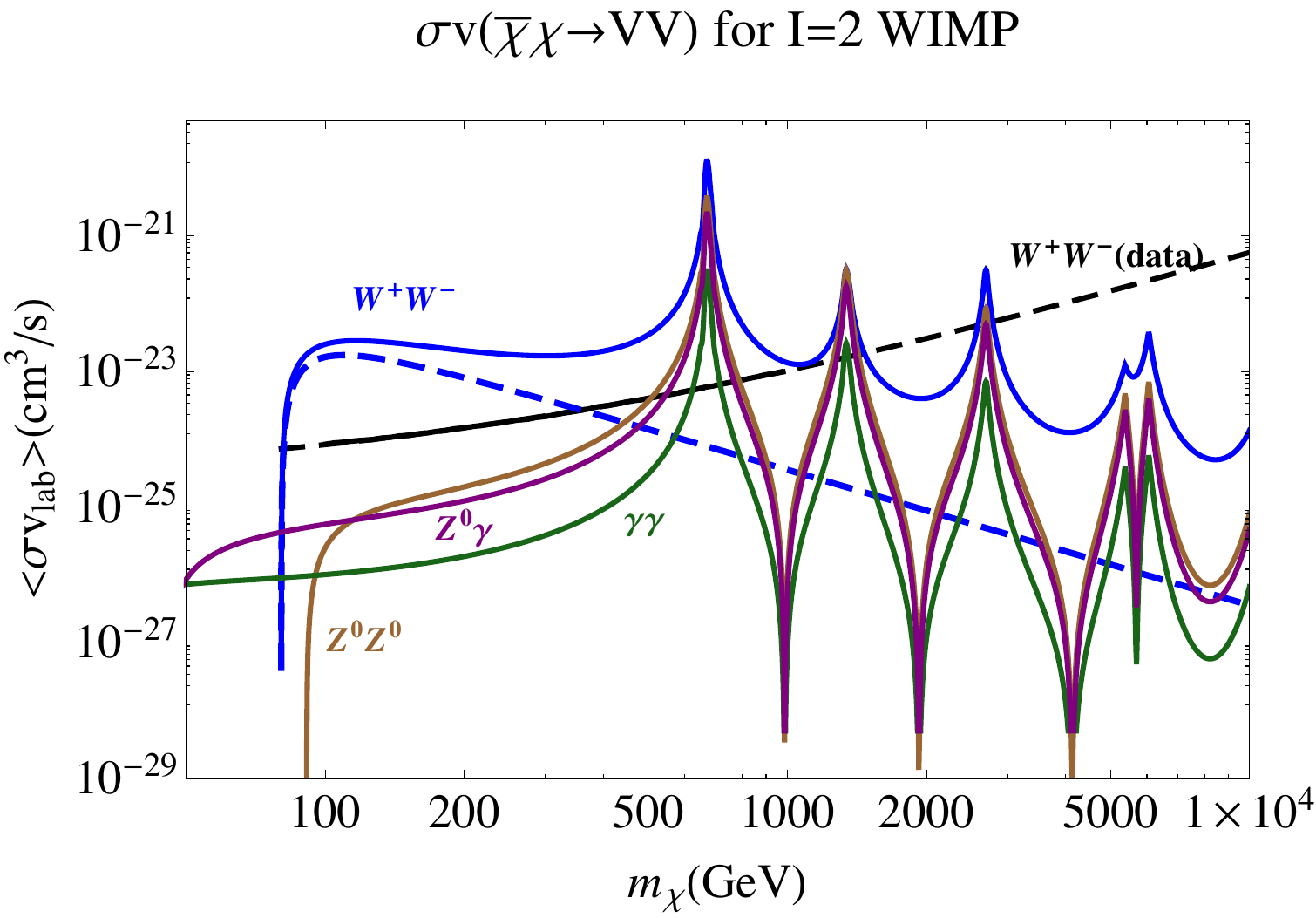}
}\subfigure[]{
  \includegraphics[width=0.45\textwidth]{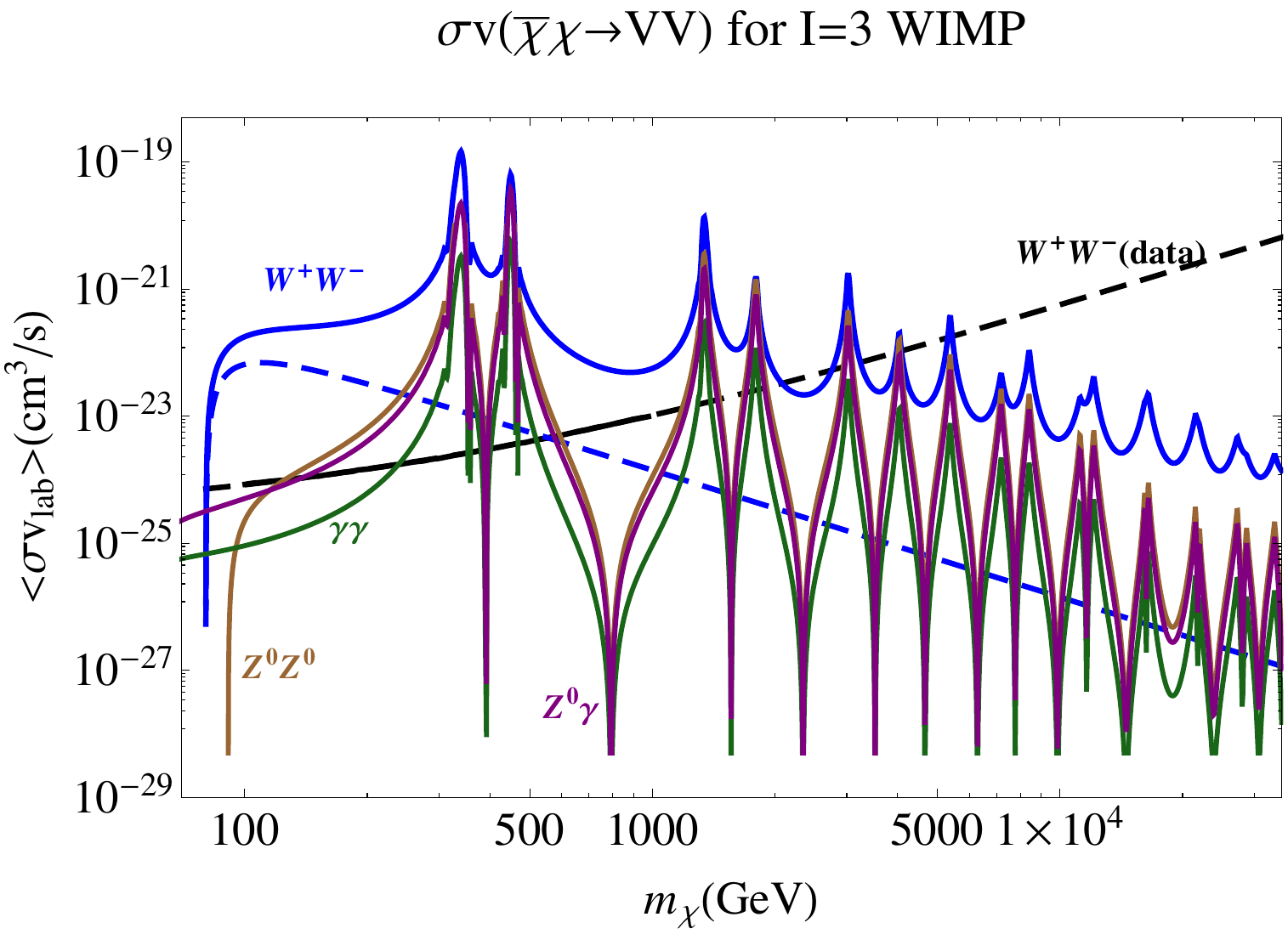}
}
\caption{(a) Predicted relic density fractions for $I=1,2,3$ compared to the data,
$\Omega_{\mathrm{nbm}}h^2=0.1187\pm0.0017$~\cite{Planck}. The solid (dashed) lines are with (without) the Sommerfeld factor.
(b) to (d):
The galactic DM annihilation cross sections for $W^+W^-, Z^0Z^0, Z^0\gamma, \gamma\gamma$ channels for different $I=1,2,3$ cases.
The solid (dashed) lines are the results with (without) the Sommerfeld factor. The $W^+W^-$ data is from\cite{FermiLat:11a} with both ends extrapolated.  }
\label{fig:WpairCS}
\end{figure}

We first give the results of case I. 
In Fig.~\ref{fig:WpairCS}(a) we show our results on relic abundance for $I=1,2,3$
and compare to the experimental  result $\Omega_{\mathrm{nbm}}h^2=0.1187\pm0.0017$~\cite{Planck}.
We take $x_f\simeq 24$ to simplify the calculations.
Solid (dashed) lines are results with (without) the Sommerfeld factor.
We see that the observed relic density can be reproduced in all three cases with TeV DM masses (see also the third column of Table~\ref{tab:result case1}).
Without the Sommerfeld factor, the masses scale as $I(I+1)$. The Sommerfeld enhancement become more prominent in the large $I$ case, and, consequently, the mass grows faster than the simple scaling.
From the figure one may easily infer that the DM masses to give correct DM relic density are larger than 50 TeV for $I>4$ and, hence, for practical purpose we shall restrict $I$ up to 3.

\begin{table}[t]
\caption{$m_\chi$ lower limits ($m_\chi^{\rm LL}$) obtained from Fermi-Lat constraints on $\chi\bar\chi\to W^+W^-$ rates, $m_\chi$ required to give correct thermal relic and the Galactic $\la\sigma v\ra$ at the corresponding dark matter masses are shown. Dark matter masses are shown in TeV, while $\la\sigma v\ra$ in cm$^3$/s. Values in parenthesis are obtained without using the Sommerfeld enhancement factors. }
\label{tab:result case1}
\begin{ruledtabular}
\begin{tabular}{l c c c c c c}
Isospin
 &$m^{\rm LL}_\chi$ (Indirect)
 & $m_\chi$ (Relic)
 & $\la\sigma v\ra(W^+W^-)$
 & $\la\sigma v\ra(Z^0Z^0)$
 & $\la\sigma v\ra(Z^0\gamma)$
 & $\la\sigma v\ra(\gamma\gamma)$
 \\
\hline $I=1$
 & $0.176$($0.166$)
 & $1.67\pm 0.01$($1.37\pm 0.01$)
 & $1.3\times10^{-24}$
 & $3.3\times10^{-25}$
 & $1.9\times10^{-25}$
 & $2.7\times10^{-26}$
 \\
$I=2$
 & $1.468$\footnotemark[1]($0.359$)
 & $7.72\pm 0.05$($4.33\pm 0.03$)
 & $1.2\times10^{-24}$
 & $1.8\times10^{-26}$
 & $1.0\times10^{-26}$
 & $1.5\times10^{-27}$
 \\
$I=3$
 & $5.446$\footnotemark[1]($0.556$)
 & $21.93^{+0.19}_{-0.08}$($8.67\pm 0.06$)
 & $1.5\times10^{-23}$
 & $2.6\times10^{-25}$
 & $1.5\times10^{-25}$
 & $2.2\times10^{-26}$
 \\
\end{tabular}
\end{ruledtabular}
\footnotetext[1]{Inferred by comparing to the extrapolated Femi-LAT data.}
\end{table}

In Fig.~\ref{fig:WpairCS}(b) to (d) we show the results of galactic $\langle\sigma v\rangle$ on WIMP annihilation for $\chi^0\bar\chi^0\to W^+W^-, Z^0Z^0,Z^0\gamma, \gamma\gamma$ channels for
WIMP candidates with different isospin ($I=1, 2, 3$) and compare them to the
milky way satellites data on the $W^+W^-$ rate~\cite{FermiLat:11a}.
We see that, when the Sommerfeld factor are removed, the $W^+W^-$ data constraints the DM masses to be heavier than few hundred GeV. However, except for $I=1$, all DM with sub-TeV mass are
excluded when the Sommerfeld enhancements are included (see also the second column of Table~\ref{tab:result case1}). The signatures of the enhancement are sizable $Z^0Z^0, Z^0\gamma, \gamma\gamma$ rates. It will be interesting to search for these processes.

In Table~\ref{tab:result case1} results on $m_\chi$ lower limits ($m_\chi^{\rm LL}$) obtained from Fermi-Lat constraints on
$\chi\bar\chi\to W^+W^-$ rates, $m_\chi$ required to give correct thermal relic
and the Galactic $\la\sigma v\ra$ at the corresponding dark matter masses are collected.
Note that these $\la \sigma v\ra$ are different from and, in fact, much than their counter part in the $x_f=24$ period as the Sommerfeld factors are more effective here.
Note that our results on $I=2$ are similar to those in \cite{MDM,Cirelli:2009uv}.

In this case we do not consider direct search as there is no data on the interesting mass regions to give the correct relic density in present and near future experiments.



\subsection*{II. $\mathbf{I=0, I_3=Y=0}$ case}

\begin{figure}[t]
\centering \subfigure[]{
  \includegraphics[width=0.47\textwidth]{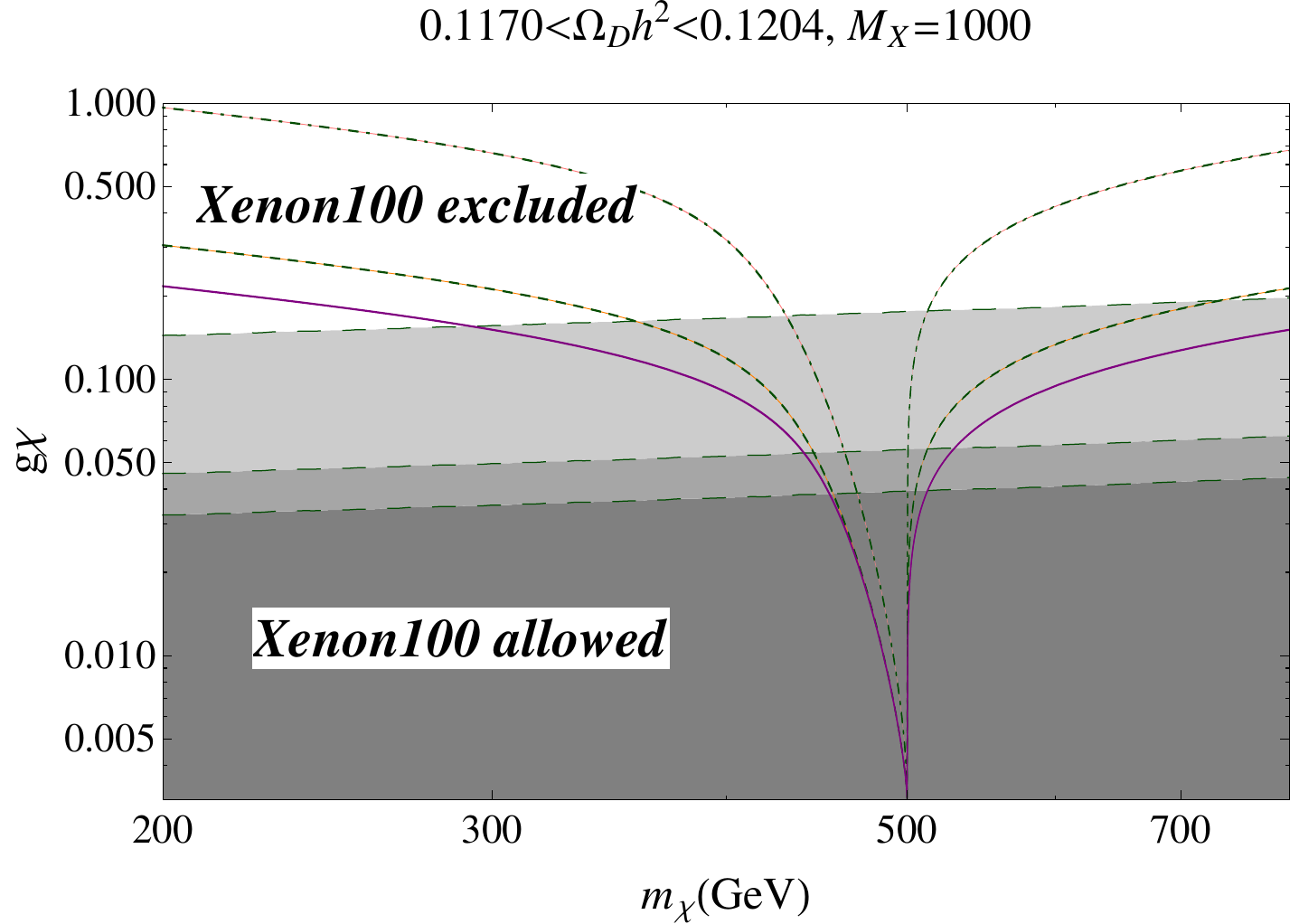}
}\subfigure[]{
  \includegraphics[width=0.47\textwidth]{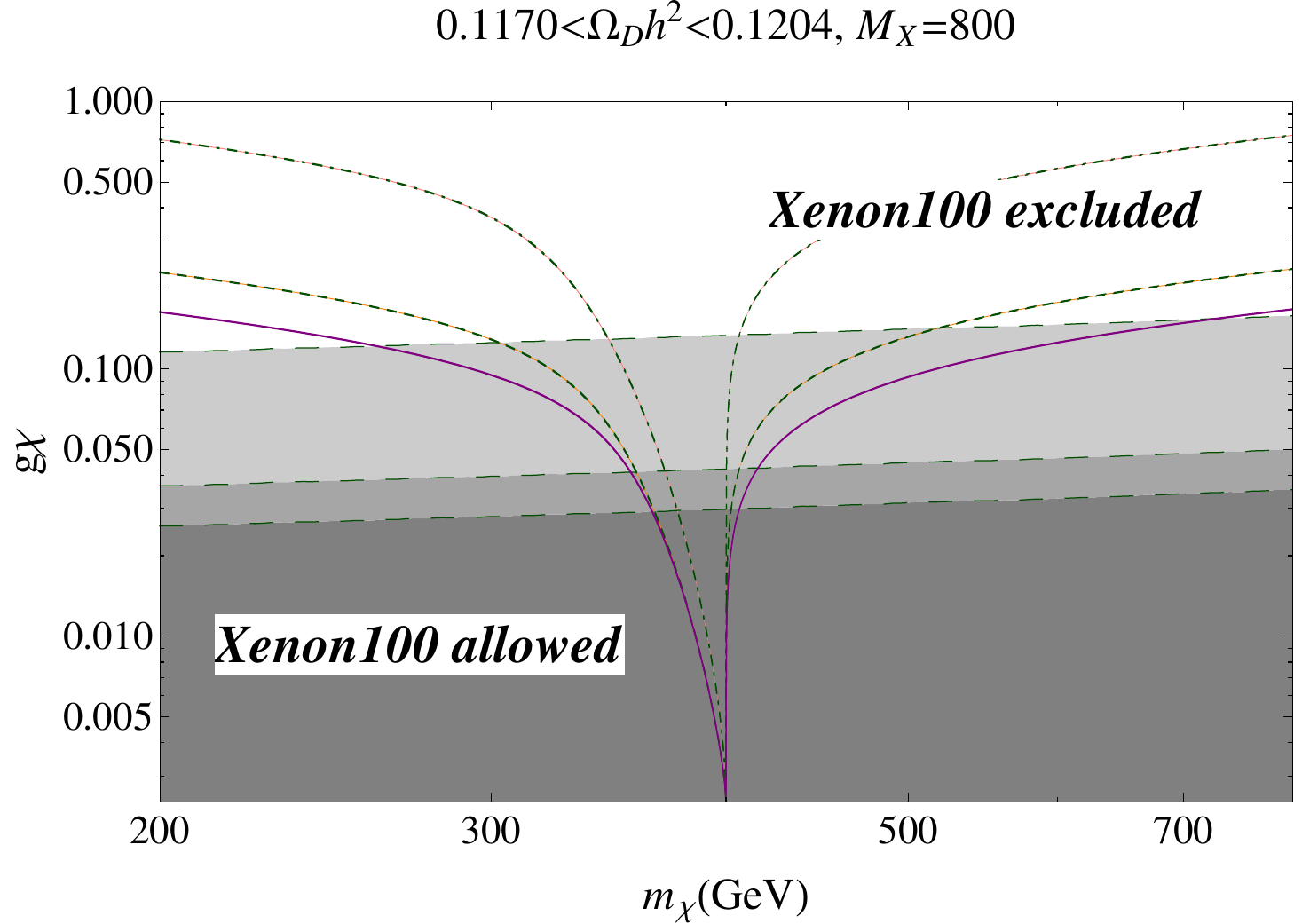}
}\\ \subfigure[]{
  \includegraphics[width=0.47\textwidth]{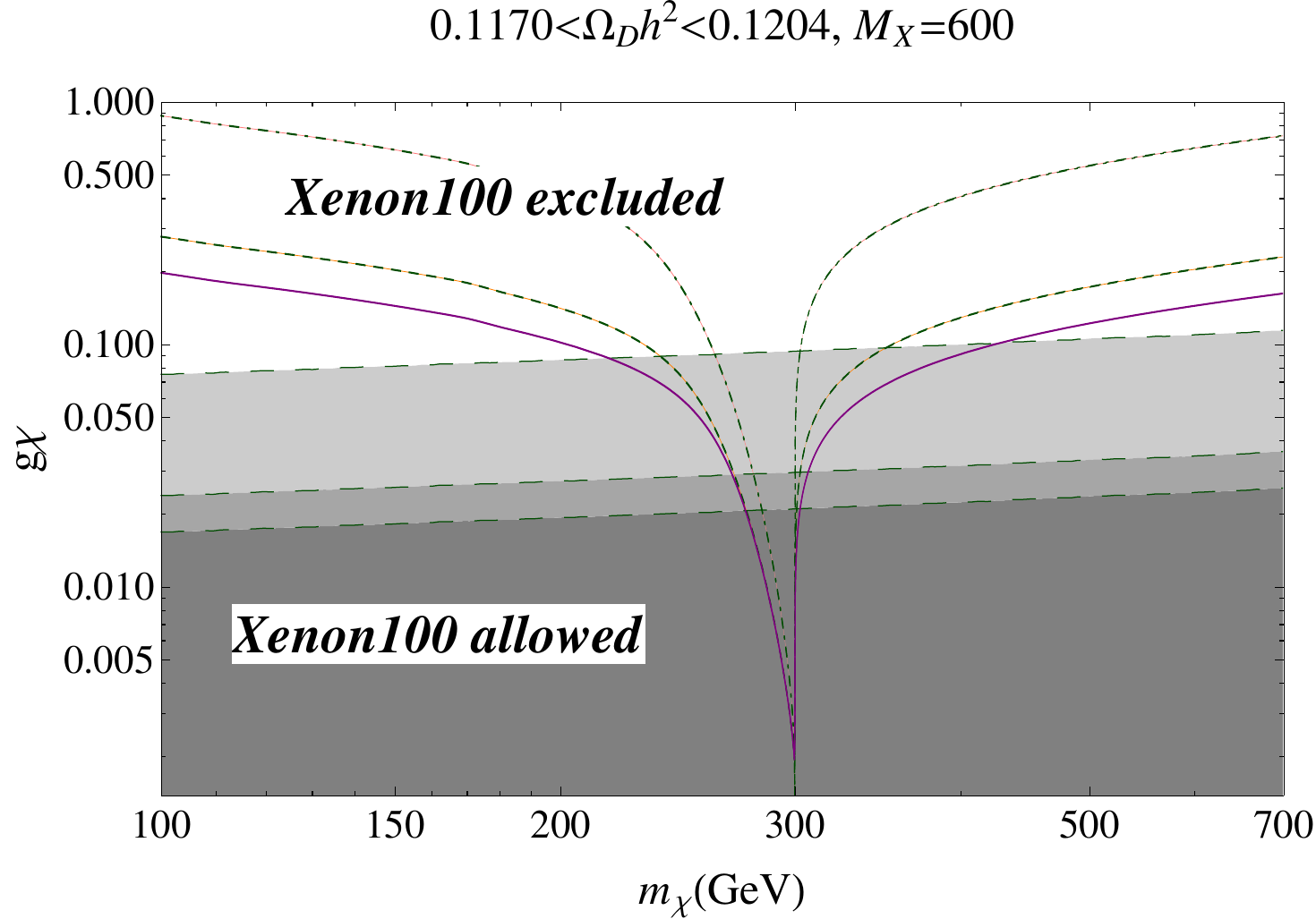}
} \subfigure[]{
  \includegraphics[width=0.49\textwidth]{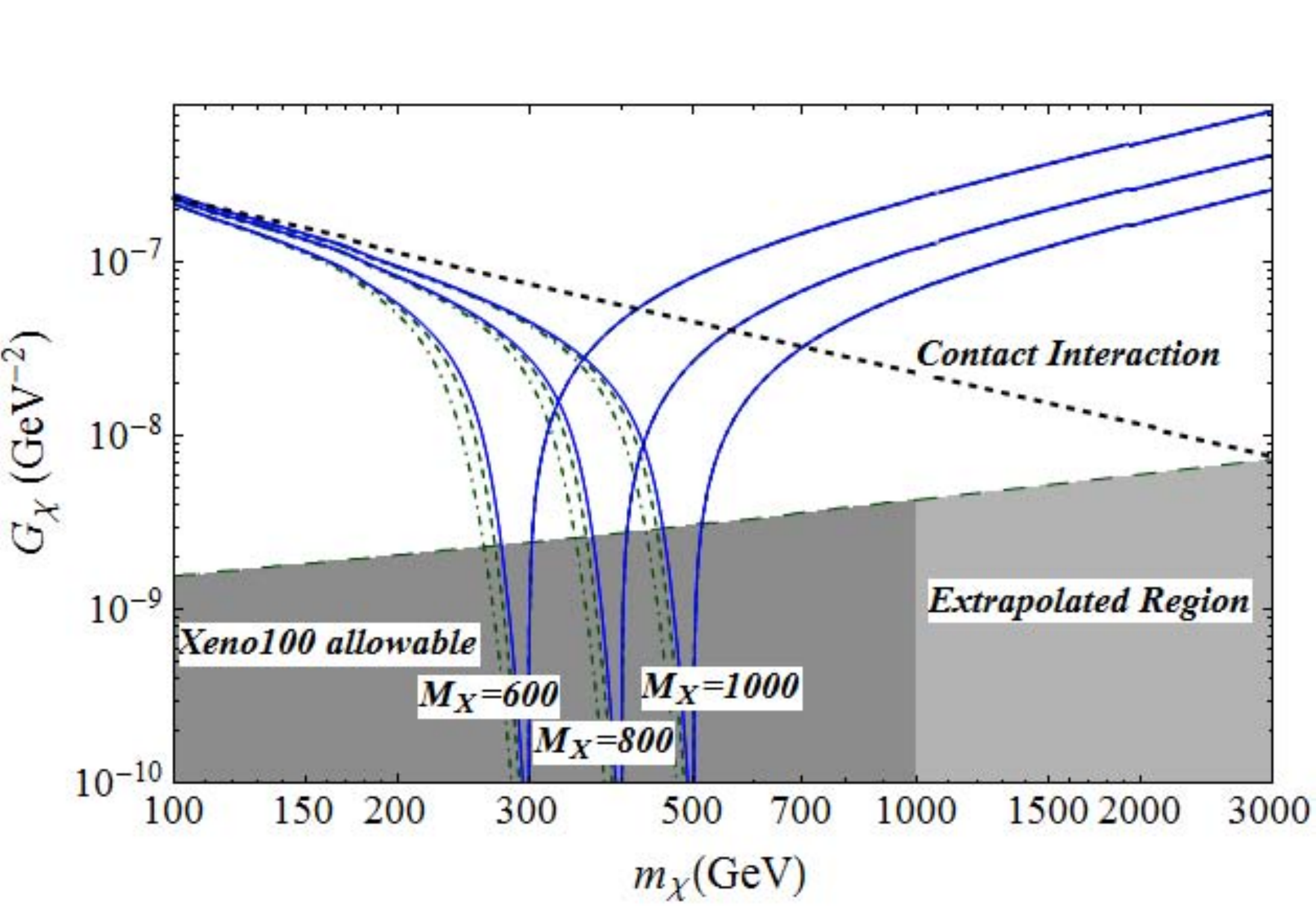}
} \caption{(a)-(c) The $g_\chi$ function for $M_X=1000,800$ and $600$ GeV
separately. In each figure, from top to bottom, we adopt
$g_f^V=0.1g_\chi$, $g_\chi$ and $2g_\chi$ to show the allowed range with
$\Omega_{\mathrm{nbm}}h^2=0.1187\pm0.0017$. The shadow region, from
top to bottom, also corresponding to $g_\chi$ constrained by Xenon100 experiment~\cite{XENON100} for different $n$ .
(d) Combining (a)-(c) results using the coupling constant $G_\chi$. The black (short dashed) line is
corresponding to contact interaction. The blue (solid) and green
(dot-dashed) lines are for containing BW resonance effect. The light
gray shadow region was obtained by extrapolating the Xenon100's
result.} \label{fig:gchi}
\end{figure}

We now turn to case II.
We shall discuss the valid parameter space first.
To simplify the numerical analysis, instead of solving Eq.~(\ref{xf}) directly,
we set the parameter value $x_f=24$, which is checked to be a good approximation,
and assume the coupling constant $g_f^V$ to be proportional to $g_\chi$
with $n\equiv g_f^V/g_\chi$. The proportionality of couplings may come from some underlying gauge symmetries,
which we will not go further into.

For given values of mediator mass $M_X$ and coupling ratio $n$, we
can solve $g_\chi$ numerically by substituting Eq.(\ref{eq:sigma
ann}) into Eq.(\ref{abundance}) and (\ref{Jfactor}). The results are
shown in Fig.~\ref{fig:gchi}. In Fig.~\ref{fig:gchi}(a)-(c), we
show the allowable range for the parameter $g_\chi$ as a function of
the DM mass with $\Omega_{\mathrm{nbm}}h^2=0.1187\pm0.0017$ for
different mediator mass, namely $M_X=1000$, $800$ and $600$ GeV. In each
figure, the curve from top to bottom, we adopt $n=0.1$, $1$ and $2$.
The shaded region are the allowed region of $g_\chi$ (with
increasing $n$ from top to bottom), which constrained by
the Xenon100 results~\cite{XENON100} of the spin-independent DM-nucleon elastic
scattering process. We note that the $g_\chi$ curve was bent down to
Xenon100 allowable region around resonance point. It is this
Breit-Wigner (BW) resonance effect that make the model to survive
from the Xenon100 experimental bound.

To further explore the physical meaning,
we define a new coupling constant $G_\chi$ such as
\begin{equation}\label{Gchi_def}
G_\chi\equiv \frac{g_\chi
g_f^V}{M_X{}^2}.
\end{equation}
In Fig.~\ref{fig:gchi}(d), we combing results in Fig.~\ref{fig:gchi}(a)-(d) using $G_\chi$.
We also plot the $G_\chi$ of the contact interaction case,
where the resonance effect is neglected and
consider only (the first term of) the contact interaction,
\begin{equation}\label{contact_L}
\mathscr{L}_{\mathrm{eff}} = -G_\chi\sum_f
\left[\bar{\chi}\gamma^\mu \chi \bar{f}\gamma_\mu f + n
\bar{f}\gamma^\mu f \bar{f}\gamma_\mu f\right],
\end{equation}
which can be obtained from Eq.~(\ref{DM_lagrangian}) by integrating
out the mediator particle $X$ when $m_X\gg m_\chi$.
It is shown in~\cite{Kingman:01}
that for an effective $eq$ contact interaction operator
$\eta_{VV}^{eq}(\bar{e}\gamma_\mu e)(\bar{q}\gamma^\mu q)$, the
effective coupling constant $\eta_{VV}^{eq}$ (which is equal to
$nG_\chi$ in our model) has an upper limit with
$nG_\chi=\eta_{VV}^{eq}\lesssim 5.01\times10^{-8}$.
This bound is comparable to the Xenon100 bound
limit $G_\chi\lesssim 3\times10^{-9}$ (see Fig.~\ref{fig:gchi}(d)).

\begin{figure}[t]
\centering \subfigure[]{
  \includegraphics[width=0.5\textwidth]{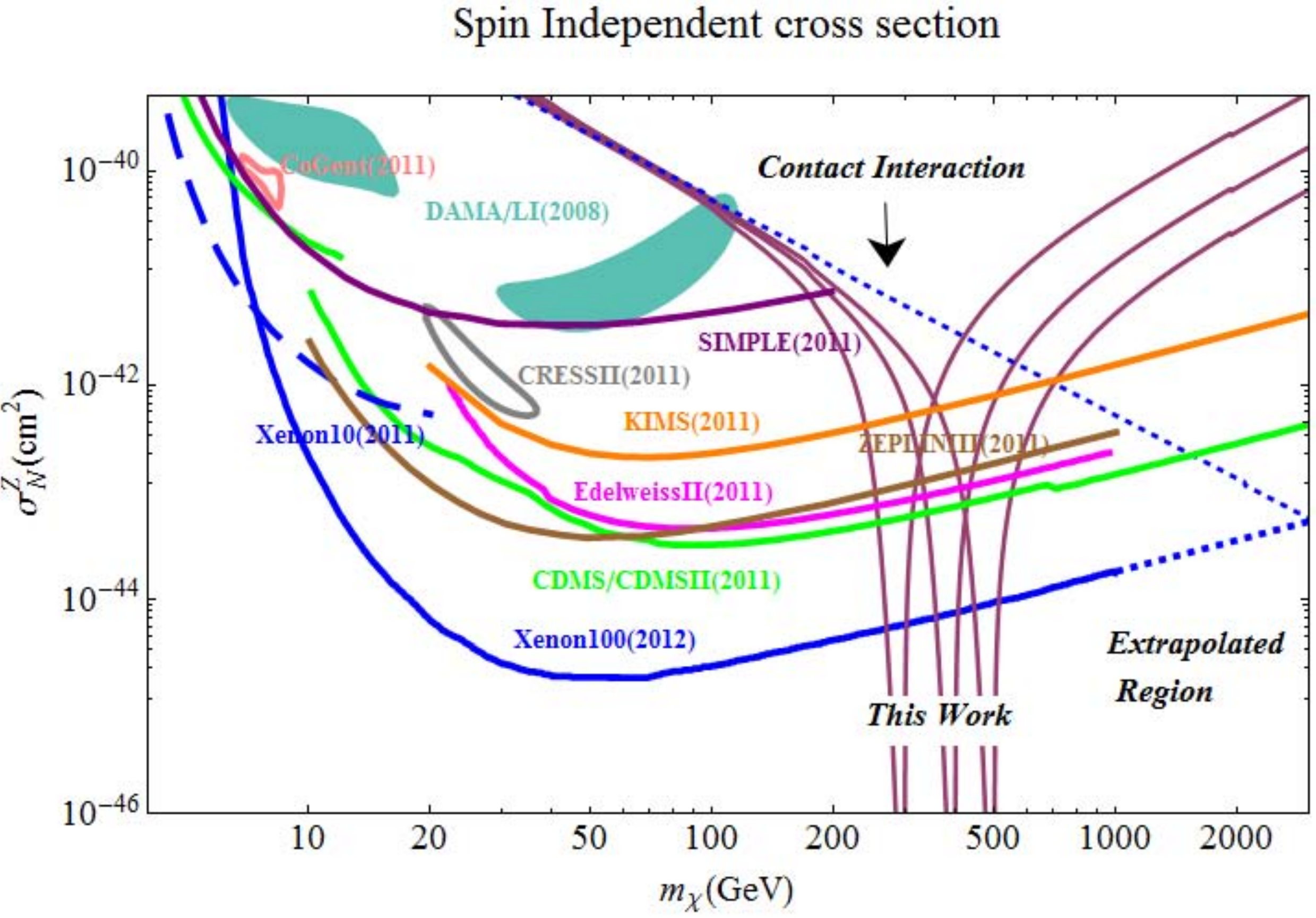}
}\subfigure[]{
  \includegraphics[width=0.5\textwidth]{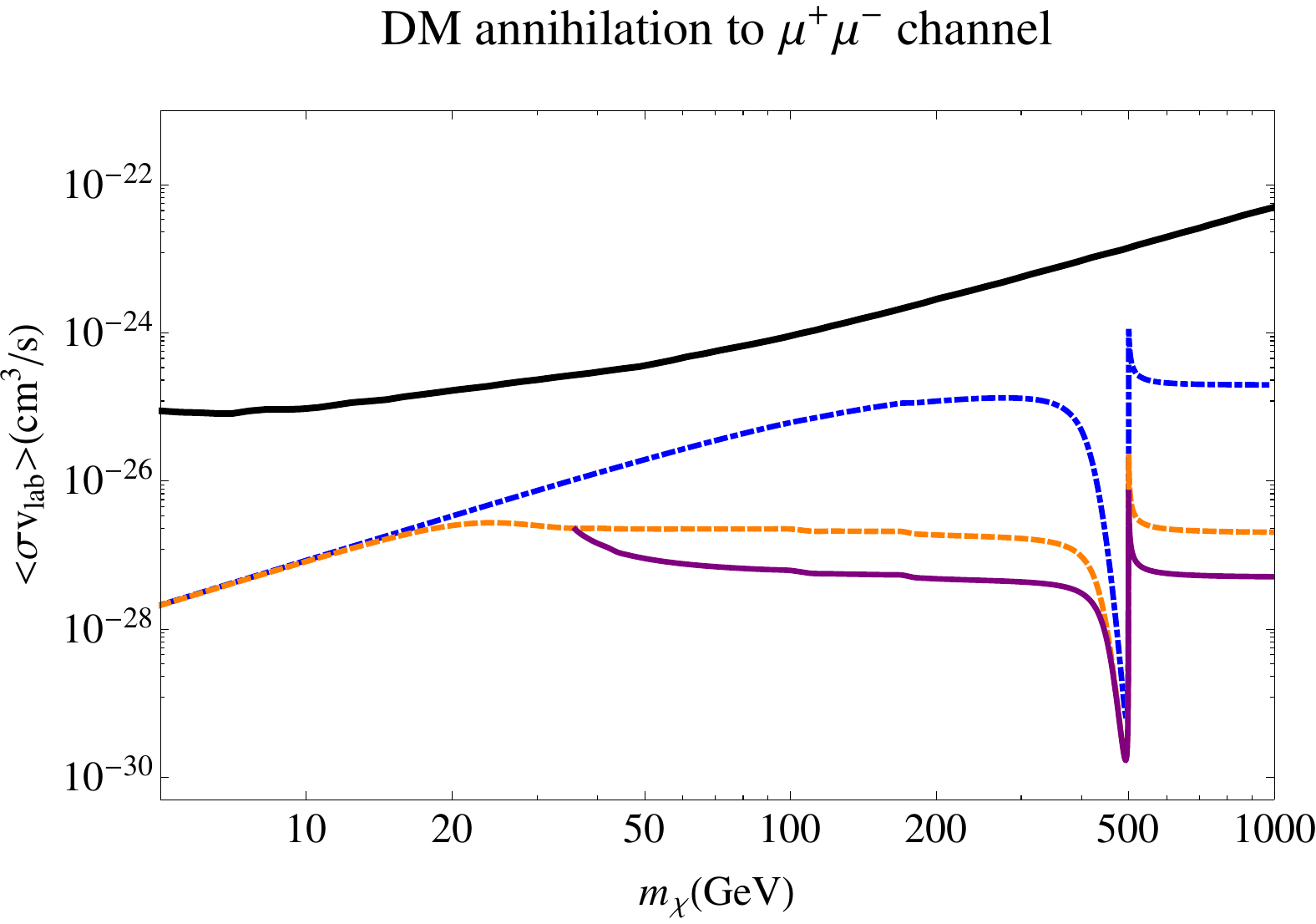}
} \caption{(a) The magnitude of elastic scattering cross section of
our model for $M_X=600, 800, 1000$ GeV with $n=1$ and direct
detection results which obtained from different experimental
group~\cite{DMtools}. We see that the BW effect indeed affect the
behavior of the curve in the resonant region. The result of contact
interaction also be showed. It reveals that the contact interaction
model can only survive at large DM mass region ($m_\chi\gtrsim 3$
TeV).(b) The DM annihilation cross section of indirect search for
the $\mu^+\mu^-$ channel. The blue(dot-dashed), orange(dashed) and
purple(solid) curve are corresponding to $g_f^V=0.1g_\chi$, $g_\chi$
and $2g_\chi$ separately with $M_X=1000$ GeV and are compared to the black solid line corresponding
to the FermiLAT result~\cite{FermiLat:11a}.} 
\label{fig:cross section}
\end{figure}

We see that, as expected, for $m_\chi\ll M_X/2$ the results (of
renormalizable interactions) are similar to the one from the
contact interaction, while for $m_\chi\gtrsim M_X/2$,
the curves are different, 
since $s$ is no longer less than $m_X^2$ (see Eq. (\ref{eq:sigma ann})).
The curve from the contact interaction can only satisfy the projected
(extrapolated) Xenon100 bound for $m_\chi\gtrsim 3$ TeV.
In other words,
the model is ruled out for $m_\chi\lesssim 3$ TeV, if the
BW effect is absent.
However, with the present of the BW effect, the model can survive from the
direct search bound with $m_\chi$ even as low as  few hundreds GeV.
Furthermore,
the allowable windows of $m_\chi$ become
broader for smaller $n$  (see
Fig.~\ref{fig:gchi}(d)).
For example, for $m_X=1000$ GeV with $n=2$, $1$ and 0.1, dark
matters having $m_\chi=500^{+12.75}_{-45.89}$,
$500^{+12.75}_{-54.58}$ and $500^{+12.75}_{-69.38}$ GeV,
respectively, can evade the direct search bound.


The corresponding typically elastic cross section $\sigma _N^Z$ for
DM and nuclei is normalized to DM-proton elastic cross section
$\sigma_p$ such that (see Appendix \ref{app:Formula})
\begin{equation}
\sigma _N^Z=\sigma _p=\frac{9\mu_p{}^2}{\pi}G_\chi{}^2
\end{equation}
with reduced mass $\mu_p=m_\chi m_p/(m_\chi+m_p)$. The result is
shown in Fig.~\ref{fig:cross section}(a). In the figure we
demonstrate the elastic scattering cross section curves of our model
for $M_X=600, 800, 1000$ GeV with $n=1$ and the contact interaction
model.
As mentioned with the resonance effect, the model can survive from the
direct search bound.

In addition to the direct search result, we also calculate the DM
annihilation cross section of indirect search for the $\mu^+\mu^-$
channel.
The result is showed in Fig.~\ref{fig:cross section}(b). The
blue(dot-dashed), orange(dashed) and purple(solid) curves are
corresponding to $M_X=1000$ GeV with $g_f^V=0.1g_\chi$, $g_\chi$ and $2g_\chi$,
respectively.

\section{Discussion and Conclusions}

\begin{figure}[t]
\centering
\includegraphics[width=0.5\textwidth]{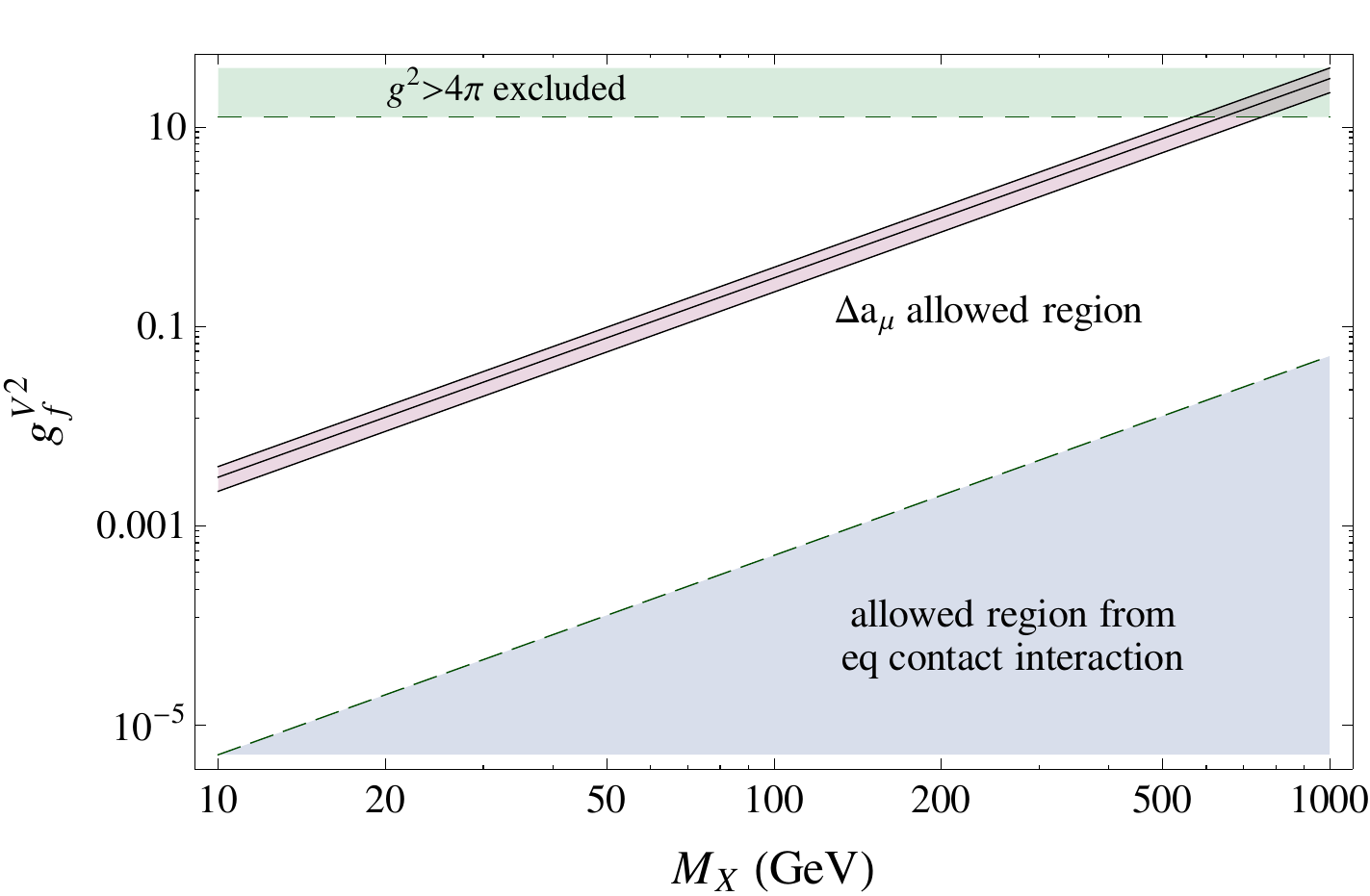}
\caption{The band corresponds to the expected parameter space to saturate $\Delta \mathrm{a}_\mu$.
The shaded region in the lower right corner is the allowed region from $eq$ contact interaction.}
\label{muon-g-2}
\end{figure}

The muon $g-2$ puzzle could be a hint for some unknown contributions
from physics beyond the SM. It will be interesting to explore the
connection with the DM sector. In Fig.~\ref{muon-g-2}, we show the
expected parameter space to saturate $\Delta a_\mu$~\cite{PDG}.
Since the expected parameter space in case II is excluded by
$(g^V_f/M_X)^2=\eta_{VV}^{eq}\lesssim
5.01\times10^{-8}$~\cite{Kingman:01}, we conclude that our model is
not sufficient to explain the deviation.

It is important to experimentally distinguish case I and case II. We note that case I has sizable Sommerfeld enhancement in  
$\chi^0\overline\chi^0\to W^+W^-, Z^0Z^0, Z^0\gamma, \gamma\gamma$ rates, which are relevant to indirect searches. For example, the $W^+W^-$ rate is enhanced by two order of magnitude from the canonical annihilation rate reaching $10^{-24}$ cm$^3$s$^{-1}$ (see Table~\ref{tab:result case1}). Fermi-LAT and iceCube will be able to search for these signatures in the future. On the other hand the indirect searches on the case II part is more or less standard. We do not expect rates to differ much from the canonical one. For example in Fig.~\ref{fig:cross section}(b), we see that the $\mu^+\mu^-$ rate is roughly $10^{-26}$ cm$^3$s$^{-1}$, as the Sommerfeld enhancement in this case is negligible.

In conclusion, we study pure weak eigenstate Dirac fermionic dark matters.
We consider WIMP with renormalizable
interaction. According to results of direct searches and the nature
of DM, 
the quantum number of DM is determined to be $I_3=Y=0$. There are only
two possible cases: either DM has non-vanishing weak isospin ($I\neq
0$) or it is an isosinglet ($I=0$). In the first case, we find that the Sommerfeld
enhancement is sizable for large $I$ DM, producing large $\chi^0\overline\chi^0\to W^+W^-, Z^0Z^0, Z^0\gamma, \gamma\gamma$ rates.
In particular, we obtain
large $\chi\bar\chi\to W^+ W^-$  cross section, which is comparable
to the latest bounds from indirect searches and
$m_\chi$ is constrained to be larger than few hundred GeV to few TeV.
It is possible to give correct relic density with $m_\chi$ higher than these lower bounds.
In the second case, to couple DM to standard model
particles, a SM-singlet vector mediator $X$ is required from
renormalizability and SM gauge quantum numbers. To satisfy the
latest bounds of direct searches and to reproduce the DM relic
density at the same time, resonant enhancement in DM annihilation
diagram is needed. Thus, the masses of DM and the mediator are
related.
Our model is not sufficient to explain the $\Delta a_\mu$ deviation.


\vskip 1.71cm {\bf Acknowledgments}

We thanks Yi-Chin Yeh and Ho-Chin Tsai for discussions.
This research was supported in part by the National Center for Theoretical Sciences
and the National Science Council
of R.O.C. under grant No NSC100-2112-M-033-001-MY3 and
NSC101-2811-M-033-016.

====================================

\appendix



\section{DM-nucleon elastic cross section}\label{app:Formula}
In this appendix, to avoid confusion, we will give some definitions
to different DM scattering cross section. The first is DM-nucleus zero
momentum transfer spin-independent(SI) cross section,
\begin{eqnarray}
\sigma^{\mathrm{SI}}_A &=& \frac{{m_\chi}^2{m_A}^2}{\pi (m_\chi +
m_A)^2}\times \frac{{g_\chi}^2}{{M_X}^4}\left[f_p Z +
f_n(A-Z)\right]{}^2
\end{eqnarray}
where $m_A$ is the mass of target nucleus with $Z$ protons and $A-Z$
neutrons and $f_{p,n}$ are the couplings to protons and neutrons
with $f_{p,n}=2g_{u,d}^V+g_{d,u}^V=3g_f^V$ in this model. By above
equation, for $Z=A=1$, we then have DM-proton cross section $\sigma
_p=9\mu_p{}^2{(g_\chi g_f^V)}^2/(\pi{M_X}^4)=9\mu_p{}^2
G_\chi{}^2/\pi$. The third is total cross section $\sigma_t=\sum_i
\eta_i \sigma_{A_i}$. Here the summation is over isotopes $A_i$ with
fractional number abundance $\eta_i$ since it is usually to include
the possibility of multiple isotopes for each detector in
laboratory. To conciliate results from different detectors, one
usually normalize the total cross section to one nucleon cross
section $\sigma_N^Z$ such as
\begin{equation}
\sigma_N^Z=\sigma_t/N = \frac{\sigma_p}{N\mu_p{}^2}\sum_i \eta_i
\mu_{A_i}{}^2 \left[Z+(A_i-Z)f_n/f_p\right]{}^2.
\end{equation}
with $N$ is normalization constant. For $f_p/f_n=1$ (isospin
symmetry), it is easy to obtain $\sigma_N^Z=\sigma_p\sum_i \eta_i
(\mu_{A_i}A_i/\mu_p)^2/N$. Because for one isotope dominated
detector, say, with proton as target, the normalized cross section
should be equal to DM-proton cross section. We then have $N=\sum_i
\eta_i (\mu_{A_i}A_i/\mu_p)^2$ and hence $\sigma_N^Z=\sigma_p$ for
different detector.


\end{document}